\documentclass[prd,amsfonts,twocolumn,nofootinbib,showpacs]{revtex4}
\RequirePackage{graphicx,  amsmath}
\begin{document}

\title{Primordial black holes from the inflating curvaton}
\author{Kazunori Kohri}
\affiliation{Cosmophysics group, Theory Center, IPNS, KEK,
and The Graduate University for Advanced Study (Sokendai),
Tsukuba 305-0801, Japan}
\author{Chia-Min Lin}
\affiliation{Department of Physics, Kobe University, Kobe 657-8501, Japan}
\author{Tomohiro Matsuda}
\affiliation{Physics Department, Lancaster University,
 Lancaster LA1 4YB, U.K.}
\affiliation{Laboratory of Physics, Saitama Institute of Technology,
Fukaya, Saitama 369-0293, Japan}
 
\begin{abstract}
\hspace*{\parindent}
The primordial black hole (PBH) formation is studied in light of the
inflating curvaton.
The typical scale of the PBH formation is determined by curvaton
 inflation, which may generate PBH with $10^{14}\mathrm{g}\le
 M_\mathrm{PBH}\le 10^{38}\mathrm{g}$ when curvaton 
 inflation gives the number of e-foldings $5\leq N_2\leq 38$. 
The non-Gaussianity of the inflating curvaton does not prevent the 
PBH formation.
\end{abstract}
\maketitle

\section{Introduction}
The origin of the structure in the Universe is due to the
primordial density perturbations that already existed when cosmological
scales started entering the horizon after the final reheating.

Inflation can generate the primary source of the cosmological
perturbation, which is usually the vacuum fluctuation of a light scalar field.
In the original inflation model the primary perturbation of the light
field (i.e, the inflaton perturbation) is
converted into the curvature perturbation at the same time when the
perturbation leaves the horizon.\footnote{This mechanism
has been used for the PBH formation in Ref.~\cite{KKSY}, 
where the secondary inflation is considered for the creation of the 
perturbation. This must be distinguished from the curvaton mechanism.}

However, later studies revealed that there are many other candidates
for the mechanism of generating the curvature perturbations after
inflation~\cite{Lyth:2002my, IRreheat, IPreheat, IPhaseTrans, Curvaton-paper,
pre-curv, Infcurvaton, Hybcurv, EKM, Pre-PBH}.  
The idea of these alternatives is very simple:
~given that there are many scalar fields in the particle physics model, 
there will be many fields displaced from their minima at the end of the
primordial inflation epoch, which leads to
 the multicomponent Universe after inflation where
any kind of isocurvature perturbations can be created.
Then the isocurvature perturbations 
can source the creation of the curvature perturbations after inflation. 

In this paper we consider an inflating curvaton~\cite{Infcurvaton,
infcurv-pre, Hybcurv, Furuuchi:2011wa, EKM} that
can lead to primordial black hole (PBH) formation.
Generation of the cosmic microwave background (CMB) perturbations is separated from the PBH formation,
since unification of these perturbations requires a highly
model-dependent argument that is not suitable for the purpose of this
paper.
Previously, such a unification scenario has been considered for the
running-mass model 
or the Type-III hilltop inflation model~\cite{Kohri:2007gq,
Kohri:2007qn,pre-papers, 
Erfani-pbh, Alabidi:2009bk, Alabidi:2012ex}.\footnote{Type-I hilltop
includes new inflation models in which the inflaton field rolls toward a larger value. Type-II hilltop includes supersymmetric hybrid inflation with a negative mass term in which the inflaton field rolls toward the origin.
Type-III hilltop includes running-mass models with a positive mass term in which the inflaton field rolls toward the origin. 
One can find classifications of these hilltop models in
Ref.\cite{Kohri:2007gq}. } 
PBH can also be produced from passive fluctuations~\cite{Lin:2012gs}.
(See also Ref.\cite{Linde:1212}.)
%added
PBH in the curvaton model has been considered in Ref.\cite{PBH-curv-pre}
in connection with the so-called ``curvaton web''
concept~\cite{Curvaton-web}.

Our strategy is to separate PBH formation from the original inflation
scenario, so that PBH formation is liberated from the CMB conditions.
Note however that the seed perturbations are generated during primordial
inflation. 

Although not mandatory, the secondary inflation may solve
cosmological problems of some specific models.
For instance:
\begin{enumerate}
\item The primordial inflation model may require e-foldings less than
      60 so that the model can explain the observed CMB.
      The short inflation could be needed to explain the spectral
      index and its running, or the inflation could be short because 
      of the fast (rapid) rolling. 
\item The model may predict unwanted relics after reheating. They must
      be diluted.
\item The model may not have an elementary dark matter (DM) candidate. Then
      the dark matter of the Universe must be explained by something
      other than the particles. 
\end{enumerate}
The most optimistic scenario of the inflating curvaton is that all these
``problems'' are solved by the secondary curvaton inflation.

In the standard curvaton mechanism, the crucial assumption is that there
are two 
components in the Universe, where one component
scales like matter ($\rho_\sigma \propto a^{-3}$)
while the other is the radiation ($\rho_r\propto a^{-4}$).
The inflating curvaton uses a slow-roll (or sometimes
fast-roll~\cite{Fast-roll}) field
instead of the oscillating 
one, which is a natural generalization of the usual curvaton
mechanism. 
We consider the slow- or fast-rolling curvaton (the inflating curvaton that
scales like $\rho_\sigma\propto a^{-3\epsilon_w}$) in the presence of
the radiation.
Then the curvature perturbation is expressed as~\cite{Lyth:2002my, Curvaton-paper} 
\begin{equation}
\zeta=(1-f)\zeta_r+f\zeta_\sigma,
\end{equation}
where
\begin{eqnarray}
f&\equiv& \frac{\dot{\rho}_\sigma}{\dot{\rho}_r+\dot{\rho}_\sigma}
=\frac{3\epsilon_w \rho_\sigma}{3\epsilon_w \rho_\sigma+4\rho_r}
\nonumber\\
\zeta_r&\equiv& -H\frac{\delta \rho_r}{\dot{\rho}_r}
= \frac{\delta \rho_r}{4 \rho_r}
\nonumber\\
\zeta_\sigma&\equiv& -H\frac{\delta \rho_\sigma}{\dot{\rho}_\sigma}
=\frac{1}{3\epsilon_w }\frac{\delta \rho_\sigma}{\rho_\sigma}.
\end{eqnarray}
Here $\epsilon_w$ is defined as
\begin{equation}
\dot{\rho}_\sigma =-3\epsilon_w H \rho_\sigma.
\end{equation}

Unlike the usual curvaton mechanism, the constancy of
$\zeta_\sigma$ requires some debate in the practical calculation, since
there is no doubt that $\zeta_\sigma$ is time dependent when the
radiation density is comparable to the curvaton density.
However, after the beginning of the curvaton inflation, 
the ratio of the radiation decreases rapidly, and $\epsilon_w$ will soon
behave like a constant.
% changed11/7
%\footnote{Although $\epsilon_w$ behave like
%constant, perturbation of $\epsilon_w$ is crucial in
%the second order calculation.
%It does not change the first order calculation considered in this
%paper~\cite{EKM}.} 
In that sense the initial quantity of $\zeta_\sigma$ must be calculated
in the inflationary phase slightly after a few e-foldings, which however
must not be later than the significant evolution of the curvature
perturbation~\cite{Infcurvaton, EKM}.
In this paper $t=t_\mathrm{ini}$ is the time when the initial quantities
are defined. 
Note that in the inflating curvaton model significant evolution occurs
when $\epsilon_w\ll 1$, while in the usual curvaton it occurs when
$\epsilon_w=1$. 
The significant evolution is accompanied by the significant change of
the parameter $r_\sigma$, which will be defined later.
The curvature perturbation cannot evolve after $r_\sigma\simeq 1$.

In this paper we are assuming that the effective curvaton potential is
always quadratic ($=\pm \frac{1}{2}m_\sigma^2\sigma^2$)
so that the ratio $\delta \sigma/\sigma$ behaves like a
constant after the horizon exit.
Thinking about the scale dependence of the perturbation,
``the initial $\delta \sigma/\sigma$ measured at the horizon exit''
depends on the corresponding scale if $\sigma$ is moving during the
primordial inflation,
and finally the distribution of $\delta \sigma/\sigma$ becomes scale dependent.
For the curvaton mechanism, a small $\sigma$ at the horizon exit
means a large perturbation $\propto \delta \sigma/\sigma$ on that
scale.\footnote{When we were finalizing 
the manuscript we found a new paper~\cite{AxionPBH}, in which the
usual curvaton mechanism has been considered for a similar PBH
formation.
The secondary inflation is the crucial difference, since it expands the PBH
scale and usually makes PBH heavier than the usual curvaton scenario.}
If the effective curvaton mass is determined by the main component of
the Universe, one can expect a small mass during
the radiation-dominated epoch~\cite{Lyth-moroi}. 
With these simple assumptions we examine the PBH formation in the
generalized curvaton mechanism.\footnote{For the scalar field $\phi$
with the bare mass $m_\phi$, the effective mass $m_{\mathrm{eff}}\sim \pm
H^2$ has been estimated~\cite{Hubble-mass} for the supergravity theory 
in the regime $H^2\gg m_{\phi}^2$. 
The possible flip of the sign has been used in supersymmetric cosmology,
for instance in Ref.\cite{AD-MSSM} for the Affleck-Dine baryogenesis.
Here the sign may depend on the main component of the energy
density (and of course on the effective interactions) at that moment. The
result is mostly due to the 
F-term contribution from the K\"{a}hler potential, which gives
$m_\phi^2\sim |F|^2/M_p^2\sim H^2$. A similar
contribution ($\sim V/M_p^2$), may appear in the
nonsupersymmetric theory of the gravity~\cite{Hubble-mass}. }

\section{Scale-dependent perturbation for the Inflating curvaton}

To show an illustrative scenario
we consider a new inflation scenario or a Type-I hilltop inflation
\cite{Kohri:2007gq} for the inflating curvaton, in which
the curvaton potential satisfies  
$\epsilon_\sigma \equiv \frac{1}{2}M_p^2 \left(\frac{V'}{V}\right)^2\ll
1$ during the curvaton inflation~\cite{Infcurvaton, Hilltop-curv}.

\begin{figure}[t]
\centering
\includegraphics[width=1.0\columnwidth]{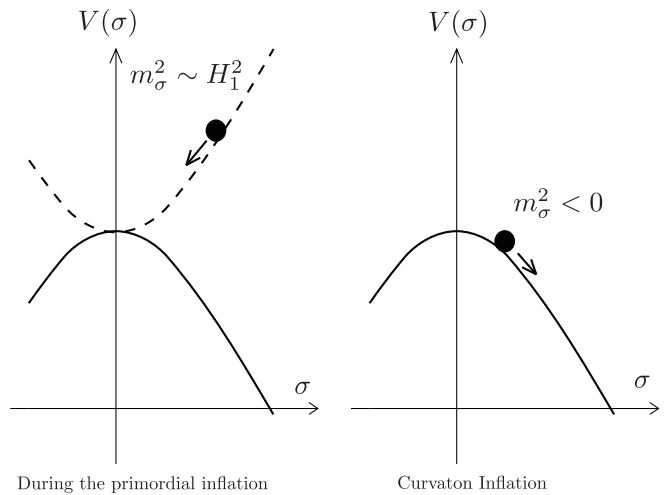}
 \caption{We are considering a scenario in which the curvaton potential
 is flipped during the primordial inflation 
because of the Hubble-induced mass, while it is negative
 during the secondary (curvaton) inflation.
The density of the curvaton is subdominant during the primordial inflation.}
\label{fig:potential}
\end{figure}

Consider the effective potential for the inflating curvaton,
\begin{equation}
V(\sigma)=V_0 + \frac{1}{2}m_\sigma(H)^2 \sigma^2,
\end{equation}
where $m_\sigma^2<0$ is possible.
Here the effective mass is determined by the main component of the Universe.
As is shown in Fig.\ref{fig:potential}, we are assuming
 $m_\sigma^2>0$ during the primordial inflation while 
$m_\sigma^2 <0$ during the curvaton inflation.
Hereafter we use the index ``$1$'' for the primordial inflation and ``$2$'' for the
curvaton inflation, respectively.

%%%% added 20130122####
In this paper we are assuming a quadratic potential for the curvaton
so that $\sigma$ and $\delta \sigma$ can share
the identical equation of motion.
At the same time the equation becomes a linear differential equation
and consequently the ratio $\delta \sigma/\sigma$ behaves like a constant.
We are avoiding any deviation from the quadratic assumption.

Let us see why the ratio can bahave like a constant when 
the quadratic potential is assumed for the curvaton.
For the simplest example, consider the equations when $m_\sigma$ and $H$ are
 constants,
\begin{eqnarray}
\label{eqomo0}
\frac{d^2\sigma}{d t^2}+3H \frac{d\sigma}{d t}+ m^2_\sigma (H)\sigma
&=&0,
\end{eqnarray}
where the equation has the solution of the
form $\sigma \propto e^{-\alpha t}$, and the coefficient $\alpha$ is given
by
\begin{equation}
\alpha\equiv \frac{3}{2}H 
\left[1\pm \sqrt{1-\left(\frac{2m_\sigma}{3H}\right)^2}\right].
\end{equation}
Defining $\beta\equiv \alpha/H$, we find
\begin{equation}
\label{typical-eq}
\sigma(t) \sim \sigma(0) e^{-\beta Ht}\equiv\sigma(0) e^{-\beta \Delta N},
\end{equation}
where $\Delta N\equiv Ht$ is the number of e-foldings spent after the horizon
exit.\footnote{Here we have omitted the subscripts, but
the evolution must be calculated for both inflationary stages.}
For the curvaton perturbation ($\sigma\equiv \bar{\sigma}+\delta
\sigma$), the equation of motion after the horizon exit
gives~\cite{Fast-roll} 
$\delta \sigma_k(t) \sim \delta\sigma_k(0) e^{-\beta\Delta N}$.
We thus find $\delta \sigma_k(t)/\sigma(t) \simeq \delta
\sigma_k(0)/\sigma(0)$ because of the cancellation of the evolution factor.
Although the mass flips its sign during the evolution, one can take the
initial condition every time at the beginning of the epoch to find a similar
evolution factor.
Here the growing solution ($\beta<0$) is possible when the mass term has
the negative sign~\cite{Hilltop-curv}. 
The equation of motion gives either a fast- or slow-rolling solution when
$|\eta_\sigma|<3/4$ (i.e, when $|m_\sigma|<\frac{3}{2}H$), 
where the eta-parameter is defined by
$\eta_\sigma \equiv \frac{m_\sigma^2}{3H^2}$.
As a consequence of the cancellation, the ratio $\delta \sigma/\sigma$
behaves like a constant as far as 
the evolution is given by the separable function. (See also Sec.\ref{mode-ind}.)

For the inflating curvaton model, the source of the
 curvature perturbation is $\delta \sigma$, which leaves the horizon during the
 primordial inflation and evolves thereafter until the curvaton
 mechanism starts to work.
At the horizon exit, the spectrum is given by
\begin{equation}
{\cal P}_{\delta \sigma_*}=\left(\frac{H_1}{2\pi}\right)^2,
\end{equation}
where $H_1$ is the Hubble parameter during the primordial inflation.

\begin{figure}[t]
\centering
\includegraphics[width=1.0\columnwidth]{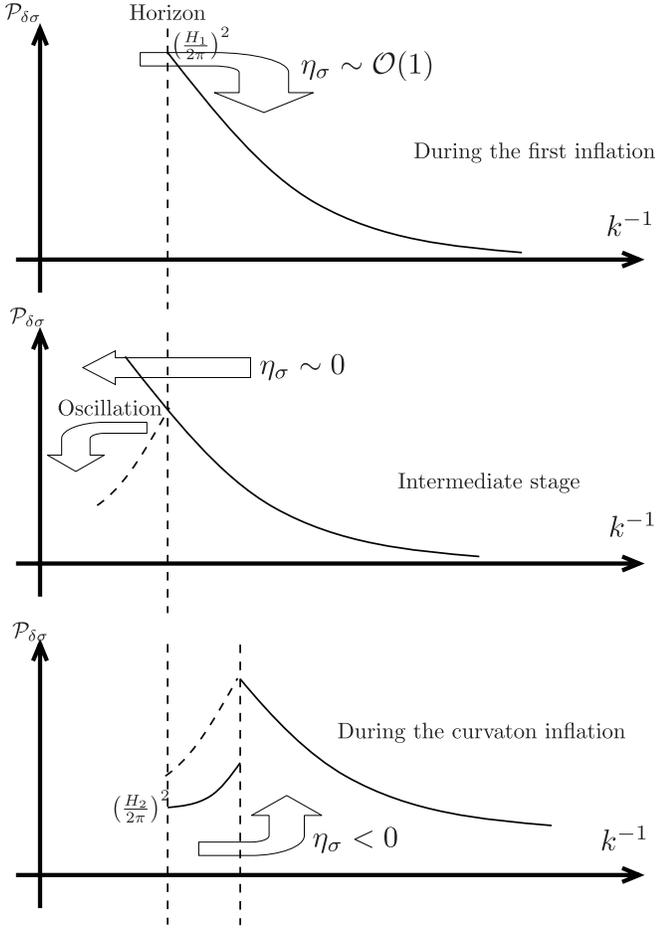}
 \caption{The field perturbation has significant scale
 dependence.
The peak appears at $k^{-1}_\mathrm{PBH}$, which corresponds to the
 scale that touches the horizon at the beginning of the curvaton inflation. 
In this picture the amplitude of the perturbation that leaves the horizon
  during the first inflation decreases during primordial inflation
 (${\cal T}_1 \ll 1$), while it increases during curvaton inflation 
(${\cal T}_2 \ge 1$).
The perturbation entering into the horizon during the intermediate
 epoch causes oscillation around its average and decreases its amplitude.
In the third picture, the perturbation newly created during the curvaton
 inflation is not dominating the amplitude at $k_\mathrm{PBH}$.}
\label{fig:peak}
\end{figure}

Since we have two inflation stages (primordial and curvaton), we also
need to consider the horizon exit during the 
curvaton inflation.
Indeed, Ref.\cite{KKSY} used the perturbation that leaves the horizon during the
second inflation and explains the small-scale 
perturbations.\footnote{Another scenario has been considered in
Refs.\cite{Kohri:2007gq, Alabidi:2009bk, Kohri:2007qn, Erfani-pbh,
Alabidi:2012ex}. 
Given that the curvature perturbation at the scale is given by 
\begin{equation}
{\cal P}_{\zeta_\varphi}\simeq\frac{{\cal P}_{\delta
 \varphi}}{2\epsilon_\varphi M_p^2}, 
\end{equation}
the significant scale dependence may appear either from $\epsilon_\varphi(k)$
(running potential) or from ${\cal P}_{\delta \varphi}(k)$.
Our paper complements the approach given in
Ref.\cite{Alabidi:2009bk} for the running $\epsilon_\varphi(k)$.}  
We are considering the opposite scenario, in which the
perturbation that exits the horizon during the primordial inflation
seeds the curvaton mechanism.
The scale dependence of the perturbation is illustrated in
Fig.\ref{fig:peak}.
We are focusing on the specific case in which the amplitude of the
curvaton perturbation becomes strongly scale dependent due to the
Hubble-induced mass during the primordial inflation.
For that reason we are not avoiding the $\eta$-problem of the
curvaton potential.

In our scenario the seed perturbation (the curvaton perturbation)
created during primordial 
inflation is highly scale dependent and it is converted into the
curvature perturbation during the curvaton inflation.
Our scenario complements the running-mass model or the Type-III hilltop
inflation model considered in Refs.~\cite{Kohri:2007gq, Kohri:2007qn,
Erfani-pbh,
Alabidi:2009bk, Alabidi:2012ex}.

For later use, we define the evolution of the perturbations as
\begin{equation}
{\cal P}^{1/2}_{\delta \sigma}(t_\mathrm{ini})
\simeq \frac{H_I}{2\pi}\times {\cal T}_1 {\cal T}_{\mathrm{int}} {\cal T}_2,
\end{equation}
where $t_\mathrm{ini}$ denotes the time when the inflating curvaton 
starts to work.
Here the $k$-dependent functions $ {\cal T}_1$, ${\cal T}_\mathrm{int}$ and $
{\cal T}_2$ denote the translation functions during primordial 
inflation (${\cal T}_1\sim e^{-\beta_1 \Delta N_1}\ll 1$ for $\beta_1>0$),
intermediate evolution (${\cal T}_\mathrm{int}\simeq 1$ for
$\beta_\mathrm{int}\sim 0$)  
and the second inflation (${\cal T}_2\sim e^{-\beta_2  \Delta N_2}> 1$
for $\beta_2<0$). 
Here $\Delta N_1$ and $\Delta N_2$ are the number of e-foldings elapsed
during each inflation stage, defined for the perturbation
on that scale ($k=k_{\mathrm{PBH}}$).
$\Delta N_1$ is defined after the perturbation leaves the horizon,
and $\Delta N_2$ is defined before the inflating curvaton starts to work at
$t=t_\mathrm{ini}$.  
${\cal T}_2$ is usually negligible for the inflating curvaton.

Here the inhomogeneity ($\delta \sigma$)
entering into the horizon can never be
kept frozen even if the potential is flat.
Oscillation around the average (not around the true vacuum) 
starts just after the horizon entry and
the amplitude is decreasing in the expanding Universe.
If the oscillation is sinusoidal, the evolution of the amplitude can be
approximated as $\propto a^{-3/2}$, which causes a significant reduction
of the amplitude, as is illustrated in the second picture of Fig.\ref{fig:peak}.
We thus find a peak on the scale that touches the horizon at the
beginning of the curvaton inflation. 
We are using this scale for the PBH formation.

For the PBH formation, we consider the spectrum of the source
perturbation ($\delta \sigma/\sigma$), focusing on the scale corresponding to the peak.
The scale of the PBH satisfies
\begin{eqnarray}
\ln\left(\frac{k_\mathrm{PBH}}{a_0 H_0}\right)&\sim&62-\Delta N_1- N_2
-\frac{1}{2}\ln\left(\frac{10^{14}\mathrm{GeV}}{H_1}\right)\nonumber\\
&&-\frac{1}{12}\ln\left(\frac{3H_1^2M_p^2}{\rho_{R1}}\right)\nonumber\\
&&-\frac{1}{12}\ln\left(\frac{3H_2^2M_p^2}{\rho_{R2}}\right).
\end{eqnarray}
Here $H_i$ is the Hubble parameter during 
inflation, and $\rho_{Ri}$ is the energy density at the reheating
after each inflation stage.
The reheating after the first inflation could be avoided, but for
simplicity we are assuming the reheating.

Since the perturbation ($k=k_\mathrm{PBH}$) 
does not enter the horizon before the curvaton
      inflation, there is a bound for $\Delta N_1$,\footnote{The number
      of e-foldings during inflation is usually given by
      $N=\ln\frac{a(t_e)H(t_e)}{a(t){H(t)}}\simeq 
\ln\frac{a(t_e)}{a(t)}$. In the above calculation the factor $1/6$
appears because $H$ varies.}
\begin{eqnarray}
\label{mindeln}
\Delta N_1 &\ge&\frac{1}{6}\ln\left(\frac{3H_1^2 M_p^2}{\rho_{R1}}\right)
+\frac{1}{4} \ln\left(\frac{\rho_{R1}}{3H_2^2 M_p^2}\right)\nonumber\\
%&=&\frac{2}{3}\ln\left(\frac{H_1}{H_2}\right)
%+\frac{1}{12} \ln\left(\frac{3H_2^2 M_p^2}{\rho_{R1}}\right)\nonumber\\
&\simeq&9 +
 \frac{2}{3}\ln\left(\frac{M_1}{10^{15}\mathrm{GeV}}\right)
-\ln\left(\frac{M_2}{10^{10}\mathrm{GeV}}\right)\nonumber\\
&&
+\frac{1}{3} \ln\left(\frac{T_{R1}}{10^{12}\mathrm{GeV}}\right),
\end{eqnarray}
where the scales are defined by $M_1^4 \equiv 3H_1^2M_p^2$ and
$M_2\equiv 3H_2^2M_p^2$.

Although the evolution of $\delta \sigma$ is highly nontrivial, 
our final result does not depend explicitly on the evolution after the
horizon exit, as far as the curvaton potential is quadratic.
That is because the ratio $\delta \sigma/\sigma$ behaves like a constant
after the horizon exit and thus the quantity evaluated just 
at the horizon exit is what we need for the PBH formation. 
Therefore, we find the perturbation using $\delta \sigma/\sigma$, which
is  evaluated at the horizon exit~\cite{Infcurvaton, EKM},
\begin{eqnarray}
\zeta&\simeq& r_\sigma \frac{2R}{3\epsilon_w}
\frac{\delta \sigma}{\sigma}\nonumber\\
&\simeq&\frac{r_\sigma A}{3\eta_2}\frac{\delta \sigma}{\sigma},
\end{eqnarray}
where the equation of motion is approximated  by
$AH\dot{\sigma}\simeq -\partial V/\partial \sigma$ with the coefficient 
$A=\beta$ for the fast-rolling and
$A=3$ for the slow-rolling.
Here $\eta_2$ denotes the value of $\eta_\sigma$ during the curvaton inflation.
The fraction
\begin{eqnarray}
r_\sigma&\equiv& \frac{3\epsilon_w \rho_\sigma}{3\epsilon_w
 \rho_\sigma+4\rho_R}\nonumber\\
&\simeq& \frac{3\epsilon_w}{3\epsilon_w +4 e^{-4N_2}}
\end{eqnarray}
is defined at the end of curvaton inflation and
throughout this paper we expect $r_\sigma\sim 1$ for simplicity.
In the above calculation we have used the ratio defined by $R\equiv \frac{m_\sigma^2\sigma^2}{2V_0}$
($m_\sigma^2<0$ is assumed during the curvaton inflation)
and the obvious relations
\begin{eqnarray}
\dot{\rho}_\sigma&\equiv&-3\epsilon_w H_2 \rho_\sigma \nonumber\\
\dot{\rho}_\sigma&\simeq& \frac{\partial V}{\partial \sigma}\dot{\sigma}.
\end{eqnarray}

The PBH is formed when the perturbation with the significant density
contrast ($\delta \rho/\rho\ge 0.1$) enters the horizon.
In our case the PBH formation occurs when the perturbation of the scale 
$\sim k_\mathrm{PBH}$ becomes accessible within the horizon.
The typical mass of the PBH is given by~\cite{Green:1999xm, Alabidi:2012ex}
\begin{equation}
\frac{k_\mathrm{PBH}}{0.00974\mathrm{Mpc}^{-1}}\simeq
\left(\frac{g_*}{g_{*eq}}\right)^{-\frac{1}{12}}
\left(\frac{M_\mathrm{PBH}}{6.67\times 10^{50}\mathrm{g}}\right)^{-\frac{1}{2}},
\end{equation}
which gives
\begin{eqnarray}
M_\mathrm{PBH}
&\simeq & 10^{46}\mathrm{g} \left(\frac{g_*}{100}\right)^{-\frac{1}{6}}
\left(\frac{k_\mathrm{PBH}}{1 \mathrm{Mpc}^{-1}}\right)^{-2},
\end{eqnarray}
where $g_*$ is the degrees of freedom in the radiation.
Let us see more details of the scenario.

First, we need to check
that $\zeta$ at the PBH scale is dominated by the curvaton.
Since the creation of $\delta \sigma$ is also possible  in the secondary 
inflation epoch, we need to compare it with the one generated during the
primordial inflation.\footnote{The spectrum of $\delta
\sigma$, which is (newly) created during curvaton inflation, must not
dominate the perturbation on the PBH scale,
since otherwise the curvaton mechanism is irrelevant for the PBH formation.}
We thus need the following condition at the beginning of the secondary
inflation:
\begin{equation}
\frac{H_2}{2\pi}<\frac{H_1}{2\pi} e^{-\beta_1 \Delta N_1},
\end{equation}
which gives a rather trivial condition,
\begin{equation}
\frac{M_2}{M_1}<e^{-\frac{\beta_1}{2} \Delta N_1}.
\end{equation}

Second, since the PBH formation requires significant density
contrast, we need 
\begin{eqnarray}
\label{PBHsecond}
{\cal P}^{1/2}_{\zeta_\sigma}
&\simeq&\frac{r_\sigma A}{3\eta_{2}}\frac{H_1}{2\pi
\sigma_{k_\mathrm{PBH}}} \sim 0.1,
\end{eqnarray}
where $\sigma_{k_\mathrm{PBH}}$ denotes the value of $\sigma$ when the perturbation
      ($k=k_\mathrm{PBH}$) leaves the horizon.
Then we find 
\begin{eqnarray}
\sigma_{k_\mathrm{PBH}} &\sim& \left(\frac{r_\sigma A}{3\eta_{2}}\right)
 \frac{H_1}{2\pi}\nonumber\\
&\sim& 10^{11}\mathrm{GeV}\left(\frac{r_\sigma A}{3\eta_{2}}\right)
\left(\frac{M_1}{10^{15}\mathrm{GeV}}\right)^2.
\end{eqnarray}

Third,
 $\zeta$ at the CMB scale must not be dominated by the curvaton.
The consistency at the CMB scale requires
\begin{eqnarray}
\label{PBHthird}
{\cal P}^{1/2}_{\zeta_\sigma}
&\simeq&\frac{r_\sigma A}{3\eta_2}\frac{H_1}{2\pi
			      \sigma_{k_\mathrm{CMB}}} <10^{-5}.
\end{eqnarray}

Therefore, from Eqs.(\ref{PBHsecond}) and (\ref{PBHthird}), we find
\begin{eqnarray}
\frac{\sigma_{k_\mathrm{PBH}}}{\sigma_{k_\mathrm{CMB}}}&\simeq&
e^{-\beta_1(N_1-\Delta N_1)}\nonumber\\
&<&10^{-4}.
\end{eqnarray}
This condition suggests that $\sigma$ moves significantly during the
primordial inflation stage.
We thus find 
\begin{equation}
N_1-\Delta N_1 >9.2 \times \frac{1}{\beta_1}.
\end{equation}
Here the scale can be given by
\begin{eqnarray}
\left(\frac{k_\mathrm{CMB}}{k_\mathrm{PBH}}\right)&\simeq& e^{N_1 -\Delta N_1},
\end{eqnarray}
where $k_\mathrm{CMB}\le 1 \mathrm{Mpc}^{-1}$.
Note that the steep potential allows significant running of the perturbation
that leads to a small $k_{PBH}$ (heavy PBH).\footnote{Note however
that Eq.(\ref{PBHthird}) gives only the upper bound for the 
perturbation. The bound gives the heaviest PBH.}

According to Carr {\it et al.} in Ref~\cite{Carr:2009jm}, only PBHs with
$10^{21} {\rm g} < M_{\rm PBH} < 10^{28} {\rm g}$ and $10^{35} {\rm g}
< M_{\rm PBH} < 10^{36} {\rm g}$ can be allowed to become dark
matter.\footnote{CMB observations might have already excluded the
latter parameter space. In addition, quite recently it was reported
that successful star formation could exclude PBH dark matter with
$10^{16} {\rm g} < M_{\rm PBH} < 10^{26} {\rm g}$~\cite{Capela:2012jz}.}
By observing $\mu$-distortion due to the dissipation of a large density
fluctuation at a small scale after the decoupling of the
double-Compton scattering, we will be able to check the amplitude of
the small-scale fluctuation by PIXIE~\cite{Chluba:2012we}.

As an illuminating example, consider $M_1\sim 10^{15}$GeV, $M_2\sim
 10^{10}$GeV and $T_{Ri}/M_i\sim10^{-3}$.
These give a typical set of the supersymmetric grand unified theory
 (SUSY-GUT) model.
Then we find that $N_1-\Delta N_1\simeq 10$ and
$N_2\sim 30$ (maximum curvaton inflation) gives
 $M_\mathrm{PBH}\simeq 10^{38}$g, or
$N_2 \sim 5$ (minimum curvaton inflation) gives
 $M_\mathrm{PBH}\simeq 10^{14}$g.
For $k_\mathrm{PBH}\sim 10^{5}\mathrm{Mpc}^{-1}$, we find 
$N_1-\Delta N_1 \simeq 11.5$ and $\beta_1> 0.8$, which allows 
$M_\mathrm{PBH}\sim 10^{36}\mathrm{g}$.

In the original argument of the inflating curvaton model,
there was a prediction for the running spectral index
($n'$)~\cite{Infcurvaton}.
(Note that this ``running'' is not for the PBH perturbation.
We are arguing here about the possible relation between the PBH mass and
the running spectral index of the CMB.)
A similar argument can be applied for the PBH model, which
suggests that a large $N_2$ (i.e, small $N_1$) gives both a large
$M_\mathrm{PBH}$ and an enhanced running of the CMB spectrum.

To be more precise about the relation between the CMB spectrum and
the PBH mass, let us assume that the primordial chaotic inflation is
driven by the 
potential $\propto \phi^\alpha$ and that it generates CMB through the
conventional curvaton mechanism. 
In that specific model we find the spectral index $n(k)-1\simeq -2\epsilon_1
=-\alpha/2N_1(k)\simeq -0.037\pm0.014$ with the running $n'\simeq
-\alpha/2N_1^2 \simeq -(0.037\pm0.014)/N_1$.
For the quadratic potential ($\alpha=2$), one will find $N_1
\sim 30$ from the spectral index, which cannot be realized without the
secondary inflation.
The running of the spectral index is enhanced because of the small
$N_1$, which could be distinguished from other predictions.
Note that in that model the second inflation is mandatory and the condition
$N_1+N_2\simeq 60-\ln(10^{-5}M_p/H_1)$ gives 
the PBH mass  $M_\mathrm{PBH}\le 10^{38}\mathrm{g}$.
Here Eq.(\ref{mindeln}), which bounds $\Delta N_1$ from below,
is given by the scale of the inflating curvaton and thus it 
can be related to the theory beyond
the standard model. 
For instance, the gravity-mediated SUSY breaking may
predict the scale of the inflating curvaton at $\sim 10^{10}$GeV
and the inflating curvaton can dilute unwanted cosmological 
relics of the model such as the gravitino or light moduli.

For the fast-roll inflation~\cite{Fast-roll}, the number of e-foldings is
usually expected to be smaller than 60 ($N_1<60$). In that sense the
secondary inflation is mandatory for the scenario.
Also, the inflating curvaton could have a very interesting application
to the rapid-roll inflation~\cite{Kofman:2007tr}. 
Using the inflating curvaton, one can increase the inflation scale of
this model to such an extent that gravity waves could be generated without
worrying about curvature perturbation.
Anyway, the curvaton-like mechanism is inevitable if
the primordial inflation may not generate the required CMB spectrum.
If the primordial inflation is short, the mechanism must be accompanied
by the secondary inflation stage, which can be used either for CMB
or for short-scale perturbations.
In that sense the inflating curvaton is quite important for these models
that do not have enough e-foldings by themselves.

There are other types of observations by which we can check the large
running. If ${\cal P}_{\zeta}$ is larger than $\gtrsim 10^{-6}$ at small
scales, ultra-compact mini halos (UCMHs) can be produced, which may be checked
by observing future lensing events or gamma rays due to dark matter
annihilation~\cite{Josan:2010vn, Bringmann:2011ut}.
The observation may help distinguish the running-mass  or
the Type-III hilltop inflation model from
the inflating curvaton model, since in the running-mass inflation model
the perturbation on the CMB scale is fixed by the usual
CMB observation and therefore its tail could be observed at smaller
scales, while there is no such bound in the inflating curvaton model.

\subsection{Model dependence}
\label{mode-ind}
Basically, the evolution of the perturbation  before the curvaton mechanism
is highly mode dependent.  
In order to avoid the model dependence,
we considered the ratio $\delta \sigma/\sigma$, for which the 
model-dependent evolution cancels when the quadratic assumption is valid.
Any deviation from the quadratic assumption can lead to a highly
mode-dependent result, which does not meet the purpose of this paper.

In our calculation the ratio $\delta
\sigma/\sigma$ evolves like a constant before the beginning of the 
curvaton mechanism.
Our observation is very simple: since $\sigma$ and $\delta\sigma$
 are sharing the same equation of motion (when the potential is
 quadratic and the perturbation is beyond the horizon), and the equation
 is a linear differential equation, their ratio
behaves like a constant.
The typical evolution is shown in Eq.(\ref{typical-eq}) for constant
$m_\sigma$ and $H$, which gives the identical evolution factor $\sim
e^{-\beta \Delta N}$ for both $\delta \sigma_k$ and $\sigma$.
The evolution is thus cancelled in the ratio.

\subsection{Non-Gaussianity}
There have been many papers suggesting that the local-type 
non-Gaussianity could prevent PBH formation~\cite{PBH-ng} when
$f_{NL}<-0.5$.
This bound could be crucial for the oscillation curvaton,
since in that model one may find $f_{NL}\simeq -1$ when $r_\sigma\simeq 1$.
On the other hand, the exact non-Gaussianity in the inflating curvaton
scenario has been calculated in Ref.\cite{EKM}, 
\begin{eqnarray}
\label{NG-eq}
f_{NL}&=&
\frac{1}{r_\sigma}
\frac{5\epsilon_w}{4R}\left(\frac{gg''}{g^{'2}}-1\right)\nonumber\\
&&-5\epsilon_w+\frac{10}{3}
+\left(\frac{5}{2}\epsilon_w-\frac{10}{3}\right)r_\sigma\nonumber\\
&&+\frac{5}{2}\epsilon_w\left(\frac{1}{r_\sigma}-\frac{1}{R}\right),
\end{eqnarray}
where the ratio is defined as
$R\equiv \pm\frac{1}{2}m_\sigma^2\sigma^2/V_0$.
Note that $R$ is negative for the hilltop potential. 
With the quadratic assumption we have 
\begin{eqnarray}
g'&=&\frac{d g}{d \sigma_*}=\frac{g}{\sigma_*}\nonumber\\
g''&=& \frac{-g+g'\sigma_*}{\sigma_*^2}=0.
\end{eqnarray}
Here the non-Gaussianity is evaluated for the PBH scale perturbation of
the inflating curvaton.

Unlike the oscillating curvaton, $r_\sigma \simeq 1$ in the inflating
 curvaton does not lead to $f_{NL}\sim -1$.
For $r_\sigma\simeq 1$ there are cancellations in Eq.(\ref{NG-eq}),
which finally gives $f_{NL}>0$ for $R<0$.
For  $r_\sigma <1$ there is no cancellation but Eq.(\ref{NG-eq}) gives 
$f_{NL}>0$.

We thus find that the sign of the non-Gaussianity parameter is plausibly
positive in our model.
The problem of the oscillating curvaton is avoided in the inflating
curvaton.

\section{Conclusion and discussion}
In this paper we have shown that the PBH formation with the typical mass
range $10^{14}\mathrm{g}\le M_\mathrm{PBH}\le 10^{38}\mathrm{g}$ is
possible if the curvaton potential is flipped due to the Hubble-induced
mass.

We have been avoiding a highly model-dependent argument, but the result
suggests that the inflating curvaton can generate the PBH that could 
be the dark matter, and/or the PBH with the interesting scale
that could be observable in future 
observations~\cite{Alabidi:2009bk, Alabidi:2012ex}.

The secondary inflation stage, which has been considered in this paper,
sometimes plays an important role in specific cosmological models.
Although it depends on the inflationary model, the spectral
 index can be tuned by changing $N_1$, which usually leads to an enhanced running
 of the spectral index.
In other cases one may expect fast-(rapid-) rolling inflaton
for the primordial inflation, which requires additional expansion.
In both cases the secondary inflation is mandatory for the scenario.

The inflating curvaton can also dilute unwanted
 cosmological relics that are produced after primordial
 inflation~\cite{second-inf}. 
In the most optimistic case, the inflating curvaton solves the above problems
 and at the same time it explains the dark matter of the Universe.

Neither artificial interaction nor fine-tuning of the potential has been
 assumed for the mechanism.

Our result also suggests that the Universe's
 many isocurvature components
may naturally lead to significant short-length perturbations.
Even if they will not source PBH formation, they might leave an observable
signal in the small-scale perturbations.

In addition, at small scales the second-order induced gravitational-wave
signal can be larger than the first order in these types
of running models~\cite{Alabidi:2012ex}.
In Ref.\cite{Alabidi:2012ex}, Alabidi {\it et al.} studied this effect
and showed that one could discriminate models by observing gravitational
waves in the future project BBO/DECIGO. 
Those observations at small scales may complement CMB observations 
at large scales, or UCMHs and $\mu$-distortion at small scales
to reveal the physics related to the PBH formation.

\section{Acknowledgment}
T.M. and K.K. thank Anupam Mazumdar for valuable discussions in the 
early stage of this work.
This work was started when T.M. was visiting Lancaster University.
C.M.L. thanks Kazuyuki Furuuchi for useful discussions.
K.K. is partly supported by the Grant-in-Aid for the Ministry of  
Education, Culture, Sports, Science and Technology, Government of
Japan Nos. 21111006, 22244030, 23540327.

\end{document}